\documentclass[sw]{agutex}
\usepackage[dvips]{graphicx}
\setkeys{Gin}{draft=false}
\authorrunninghead{MURRAY ET AL.}

\titlerunninghead{THERMOSPHERIC MODELLING DURING SOLAR CYCLES 23 AND 24.}

\authoraddr{Corresponding author: S. A. Murray,
Met Office, FitzRoy Rd., Exeter, Devon, EX1 3PB, UK
(sophie.murray@metoffice.gov.uk)}

\begin{document}

%% ------------------------------------------------------------------------ %%
%
%  TITLE
%
%% ------------------------------------------------------------------------ %%

\title{Assessing the performance of thermospheric modelling with data assimilation throughout solar cycles 23 and 24.}
%
% e.g., \title{Terrestrial ring current:
% Origin, formation, and decay $\alpha\beta\Gamma\Delta$}
%

%% ------------------------------------------------------------------------ %%
%
%  AUTHORS AND AFFILIATIONS
%
%% ------------------------------------------------------------------------ %%

%Use \author{\altaffilmark{}} and \altaffiltext{}

% \altaffilmark will produce footnote;
% matching \altaffiltext will appear at bottom of page.

\authors{S. A. Murray,\altaffilmark{1}
 E. M. Henley,\altaffilmark{1} D. R. Jackson\altaffilmark{1}, and S. L. Bruinsma\altaffilmark{2}

\

\emph{Accepted for publication in Space Weather.}} 

\altaffiltext{1}{Met Office, Exeter, Devon, UK.}

\altaffiltext{2}{CNES, Toulouse, France.}

%% ------------------------------------------------------------------------ %%
%
%  KEY POINTS
%
%% ------------------------------------------------------------------------ %%

%Data assimilation systems developed for semi-empirical and physical thermospheric models.
%Model results compared to satellite observations at solar maximum and solar minimum.
%A mean improvement of \sim$ 4\% was found using data assimilation with the general circulation model.

%% ------------------------------------------------------------------------ %%
%
%  ABSTRACT
%
%% ------------------------------------------------------------------------ %%

% >> Do NOT include any \begin...\end commands within
% >> the body of the abstract.

\begin{abstract}
Data assimilation procedures have been developed for thermospheric models using satellite density measurements as part of the EU Framework Package 7 ATMOP Project. Two models were studied; one a general circulation model, TIEGCM, and the other a semi-empirical drag temperature model, DTM. Results of runs using data assimilation with these models were compared with independent density observations from CHAMP and GRACE satellites throughout solar cycles 23 and 24. Time periods of 60 days were examined at solar minimum and maximum, including the 2003 Hallowe'en storms. The differences between the physical and the semi-empirical models have been characterised. Results indicate that both models tend to show similar behaviour; underestimating densities at solar maximum, and overestimating them at solar minimum. DTM performed better at solar minimum, with both models less accurate at solar maximum. A mean improvement of $\sim$ 4\% was found using data assimilation with TIEGCM. With further improvements, the use of general circulation models in operational space weather forecasting (in addition to empirical methods currently used) is plausible. Future work will allow near-real-time assimilation of thermospheric data for improved forecasting. 
\end{abstract}

%% ------------------------------------------------------------------------ %%
%
%  BEGIN ARTICLE
%
%% ------------------------------------------------------------------------ %%

% The body of the article must start with a \begin{article} command
%
% \end{article} must follow the references section, before the figures
%  and tables.

\begin{article}

%% ------------------------------------------------------------------------ %%
%
%  TEXT
%
%% ------------------------------------------------------------------------ %%

\section{Introduction}
\label{intro}

The Advanced Thermosphere Modelling of Orbital Prediction (ATMOP) project was designed to provide a European capability for nowcasting and forecasting of the thermosphere. Changes in thermospheric density affect the drag experienced by low-orbiting satellites and hence their orbits. The resulting loss in orbit predictability is problematic for space agencies and satellite operators, with thousands of objects orbiting the Earth and the majority of them in low-earth orbit \citep[see review by][]{vallado14}. Space weather is one cause for these density changes, as the associated changes in solar or geomagnetic activity affects the coupled thermosphere-ionosphere system \citep{rees89,huba14}.

A major focus of ATMOP was to investigate the benefit of data assimilation to predictive models. While this is a standard method in meteorological forecasting, it is a relatively new approach within the space weather community. This paper compares both semi-empirical and physical models using this new approach. Semi-empirical models are more rapid to run, and typically outperform physical models, hence are currently used for operational orbit computations. However, the physical models may perform better in unusual conditions, such as those encountered in geomagnetic storms. Two models are the focus of this work - the first a version of the semi-empirical Drag Temperature Model \citep[DTM-2012;][]{bruinsma03}, and the second the physical Thermosphere Ionosphere Electrodynamics General Circulation Model \citep[TIEGCM;][]{richmond92}. 

A large amount of previous work has examined the relative accuracy of thermospheric general circulation models compared to empirical methods. For example, \citet{anderson98} intercompared five models including TIEGCM, in order to determine why several physical models consistently underestimate the ionospheric F region peak electron density. \citet{qian12} examined temporal and spatial variations of thermospheric density via a comparison of TIEGCM with an empirical model, NRLMSISE-00 \citep{picone02}. In-situ observations from satellites are used in these comparisons to determine whether the models accurately represent thermospheric conditions \citep[see also,][]{buonsanto97,pavlov00,forbes05,sutton05,tsagouri13}. Readily available satellite data are vital for space weather operations, being used in data assimilation techniques to improve the thermospheric models that are used to predict the drag experienced by low-orbiting spacecraft \citep{rowell06}. 

A typical atmospheric model set up with boundary conditions closely matching reality will nevertheless produce predictions which gradually diverge from reality due to incomplete characterisation of the physics and chaotic effects. Thermospheric models are also strongly influenced by external drivers; geomagnetic and solar drivers (which are generally represented by way of input indices to the models), and the lower and upper boundaries implemented to represent the interaction with the lower atmosphere below and plasmasphere above. By assimilating observations into a physical model at a given time, model fields at that time can be brought closer to the real data values. Bringing the model state closer to reality reduces the divergence of the ensuing prediction from reality, thereby increasing the accuracy of the model forecast. Data assimilation repeats this process regularly, to provide a running forecast with better overall accuracy.

Much previous thermospheric data assimilation work has focused on using Kalman filters \citep{rowell04}, particularly the development of ensemble Kalman filtering methods \citep[EnKF;][]{codrescu04,matsuo12} with recent advances in computer technology. \citet{matsuo12} found the root-mean-square (RMS) error of model-predicted neutral mass density could be reduced by up to 50\% in the case that as many as nine ensemble members are used. \cite{codrescu04} found the error of their model initial state reduced from $\sim$25\% to $\sim$10\% using an EnKF technique with ten members. \citet{matsuo13} used EnKF with inferred neutral density satellite observations to improve a model neutral density specification in the vicinity of the satellite. \citeauthor{matsuo13} found that a global impact was achieved if accompanied by the estimation of the primary driver of the density variable (such as solar EUV flux), highlighting the importance of external forcing to the system.

Two data assimilation methods were developed in ATMOP; one using DTM-2012, and the other TIEGCM. A version of DTM was developed that can assimilate total density data in near-real time, which will make orbit predictions significantly more accurate. Data assimilation techniques were also developed for use with TIEGCM with the aim to create a more physically accurate global analysis and forecast system for the thermosphere. While the assimilation methods developed are described in detail elsewhere \citep[see][and \emph{Henley et al}, manuscript in preparation]{bruinsma12}, this paper provides first results of an intercomparison of the newly developed models. Both TIEGCM and DTM-2012 are compared with observations taken at various periods throughout solar cycles 23 and 24 in order to determine if a more complex physical model with data assimilation could be more accurate for forecasting efforts than the currently-used empirical methods. This validation effort is crucial to determine if these new techniques are beneficial for current space weather operations. The observations and models used for the comparison will be outlined in Section~\ref{obs}, and methods used for analysis described in Section~\ref{method}. The results of the study will be presented in Section~\ref{results}, while some conclusions of the findings and possible future work will be discussed in Section~\ref{concl}.

%------------------------------------

\section{Observations and Models}
\label{obs}

Atmospheric drag on satellites varies strongly as a function of thermospheric mass density \citep[for further discussion see][]{qian12,vallado14}. This work uses in-situ observations of atmospheric densities in the upper thermosphere for both the data assimilation techniques and the validation and intercomparison of the model results. The thermospheric densities were inferred from accelerometers that fly on low-earth-orbiting satellites, namely the CHAllenging Minisatellite Payload \citep[CHAMP; ][]{wickert01} and the Gravity Recovery And Climate Experiment \citep[GRACE; ][]{tapley04}. See \citet{bruinsma04} for specific details on how the densities are calculated. The height at which the densities are available depends on the satellite orbital height; this is $\sim$450--300~km for CHAMP and $\sim$490--410~km for GRACE (as of 2014 October). The height value is dependant on the time period of observations; CHAMP data are available from 2000 to 2010, and GRACE data from 2002 to present. The satellites observe at $\sim$15 orbits per day at all $360^\circ$ of longitude, with a spacing of $25^\circ$ at the equator per day, and a latitude resolution of 0.5--1$^\circ$. The horizontal resolution is 80 and 40~km along the orbit for CHAMP and GRACE, respectively. Observations are made every 10~s with CHAMP and 5~s with GRACE. The satellite data have a mean and RMS observation error both ranging between 1--10\%, which is a function of latitude and geomagnetic activity. 

The drag temperature model used for comparison in this paper, DTM-2012 (hereafter DTM), is an improved version of DTM-2009 \citep{bruinsma12} that is trained on a larger data set. It has been constructed using the full CHAMP and GRACE high-resolution accelerometer inferred density data sets, besides new mean total density data as well as historical mass spectrometer data. The 81-day mean and 1-day delayed solar radio flux at 10.7cm (F10.7) are used as solar inputs, and the A$_\mathrm{m}$ geomagnetic index is also used. DTM is constructed by fitting to the underlying density database, as good as possible in the least-squares sense, to reproduce the mean climatology of the thermosphere \citep[see][]{bruinsma04}. The result includes a density value at a particular latitude and longitude between 200--1000~km, with up to $1^\circ$ resolution. Errors on short time scales of a few days or less can be of the order of tens of percent. A more detailed description of the updated DTM with data assimilation can be found in \citet{bruinsma12}.

The physical model used in this work, TIEGCM, is a first-principles, three-dimensional, non-linear representation of the coupled thermosphere and ionosphere system. TIEGCM uses spherical geographic coordinates, with latitude ranging from $-87.5^\circ$ to $87.5^\circ$ in $5^\circ$ increments, longitude ranging from $-180^\circ$ to $180^\circ$ in $5^\circ$ increments, and a 60-second time step. The lower boundary at $\sim$97~km extends up to $\sim$500--700~km depending on solar activity. The migrating and semi-diurnal tides are specified at the lower boundary using the Global Scale Wave Model \citep{hagan95}, which does not consider the effects of planetary waves and non-migrating tides. The vertical coordinate is a log-pressure scale, ln(p$_\mathrm{0}$/p), where p is pressure and p$_\mathrm{0}$ is a reference pressure, and pressure levels range from $-7$ to $7$ increasing in half-scale-height increments. 

Neutral density observations were assimilated into the model using an ensemble optimal interpolation method \citep[EnOI; ][]{oke02, evensen03}. EnOI uses an ensemble approach to help determine how much trust to put in the model prediction at any given time (as opposed to how much trust to put in the observations). An ensemble of nine models are run offline, with each model regularly perturbed with smoothed random temperature fluctuations. A background density error covariance matrix is then calculated from these non-assimilative, independent ensemble members.  This is combined with the background and observations to produce a density analysis. The analysis is then converted to a temperature analysis (as density is a derived field in TIEGCM), which is used to initialise the model run. This is the first time an EnOI method has been used for thermospheric data assimilation, with much previous work focusing on EnKF (see Section~\ref{intro}). For further details on the assimilation scheme, see Henley et al$.$ (manuscript in preparation, 2015).

For consistency with DTM input parameters, TIEGCM was run using the Heelis high-latitude ion convection model \citep{heelis82} as a magnetospheric input, which uses F10.7 solar flux and K$_\mathrm{p}$ planetary geomagnetic index as inputs. The model provides various advantages for development, speed, robustness, and scientific return compared to empirical modelling. Results from TIEGCM have been previously compared favourably with CHAMP data on a large scale \citep{sutton08, qian12}, although it has also been noted that TIEGCM RMS errors typically gradually increase with a decline in solar activity \citep{kim11}.

%------------------------------------

\section{Comparison}
\label{method}

TIEGCM and DTM results were compared with CHAMP and GRACE observations at three periods throughout solar cycles 23 and 24. Solar equinox periods were selected for ease of comparison of model-run results. For solar minimum, a 60-day period was chosen beginning 2009 March 15, and for solar maximum another 60-day period beginning 2003 March 15 was selected. To examine the impact of more severe stormy conditions, a final 60-day period was examined beginning 2003 October 15, which encompasses the Hallowe'en storms. These storms are defined as a sequence of events between 2003 October 28 to November 4, during which massive solar flares and geomagnetic storms occurred. The impact of this event on the near-Earth environment has been heavily studied, including analysis of observations \citep[e.g.,][]{thomson04,mannucci05,bruinsma06} as well as the use of simulations \citep[e.g.,][]{toth07,manchester08}. \citet{gopalswamy05} gives a detailed review of the events, while \citet{oler04} outlines the prediction performance of space weather forecasting centres worldwide following these events.

A large database of observations were assimilated into DTM (see Section~\ref{obs}) before running the model on the time periods selected. TIEGCM was run with 1-hourly data assimilation using the EnOI system (see Section~\ref{obs}) along with CHAMP observations for these periods. It is worth noting the impact of assimilation is at a local level since the satellite data give density values at a particular latitude, longitude, and height, and a localisation scheme is also applied. The CHAMP data are decimated from their original 10~s cadence to 120~s cadences in order to help stability, and this also provides semi-independent CHAMP data sets to use for validation. More formal validation is provided by independant GRACE observations.

The model results were interpolated to the altitude, latitude and longitude of the particular spacecraft at that time for comparison purposes. The interpolation procedure used was the same as that used within the TIEGCM data assimilation code for consistency (Henley et al., manuscript in preparation, 2015).  In order to accurately compare the output of the two models, DTM was run for the particular altitude, latitude, and longitude of the spacecraft at time of observation. DTM densities were calculated every 5 or 10 seconds (depending whether CHAMP or GRACE was being compared), while TIEGCM resulting densities were saved every 15 minutes. It is worth noting that both CHAMP and GRACE spacecraft have different local times for the periods studied here, the difference ranging between $\sim$1--2~LT. It useful to compare the model results to these observations for validation purposes since they have an altitude difference of $\sim$100--150km for the time periods studied. However, a more in-depth comparison of latitudinal or longitudinal variation is not undertaken here.

\begin{figure}[t]
\centering
\noindent\includegraphics[width=\columnwidth]{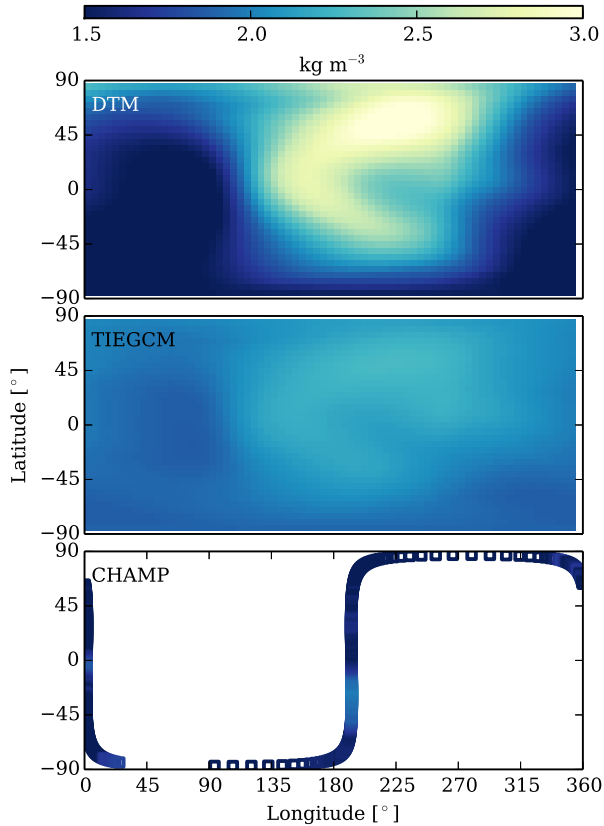}
\caption{DTM (upper row) and TIEGCM (middle row) density values on 2009 March 01 00:00UT, and CHAMP (lower row) density values between 2009 March 01 00:00--01:30UT. The DTM and CHAMP values have been interpolated to TIEGCM pressure level 21 at $\sim 10^{-6}$~Pa ($\sim$335km).}
\label{figure_context}
\end{figure}

Typical model outputs are presented in Figure~\ref{figure_context}. A 2D map of DTM density at 2009 March 01 00:00UT is shown in the upper row, with an equivalent TIEGCM map in the middle row. CHAMP densities are also plotted in the lower row for an entire 90-minute orbit. Note the CHAMP spacecraft was at an altitude of $\sim$328km during this orbit, however the DTM and CHAMP densities have been interpolated to TIEGCM pressure level 21 ($1.18\times10^{-6}$~Pa) for ease of comparison. TIEGCM tends to have a narrower range of density values compared to DTM. It is also worth noting that both models overestimate the density in the points corresponding to the CHAMP measurements. This is a result that is emulated throughout solar minimum, as will be discussed in the following section. 

%------------------------------------

\section{Results}
\label{results}
\begin{figure*}
\centering
\noindent\includegraphics[width=\textwidth]{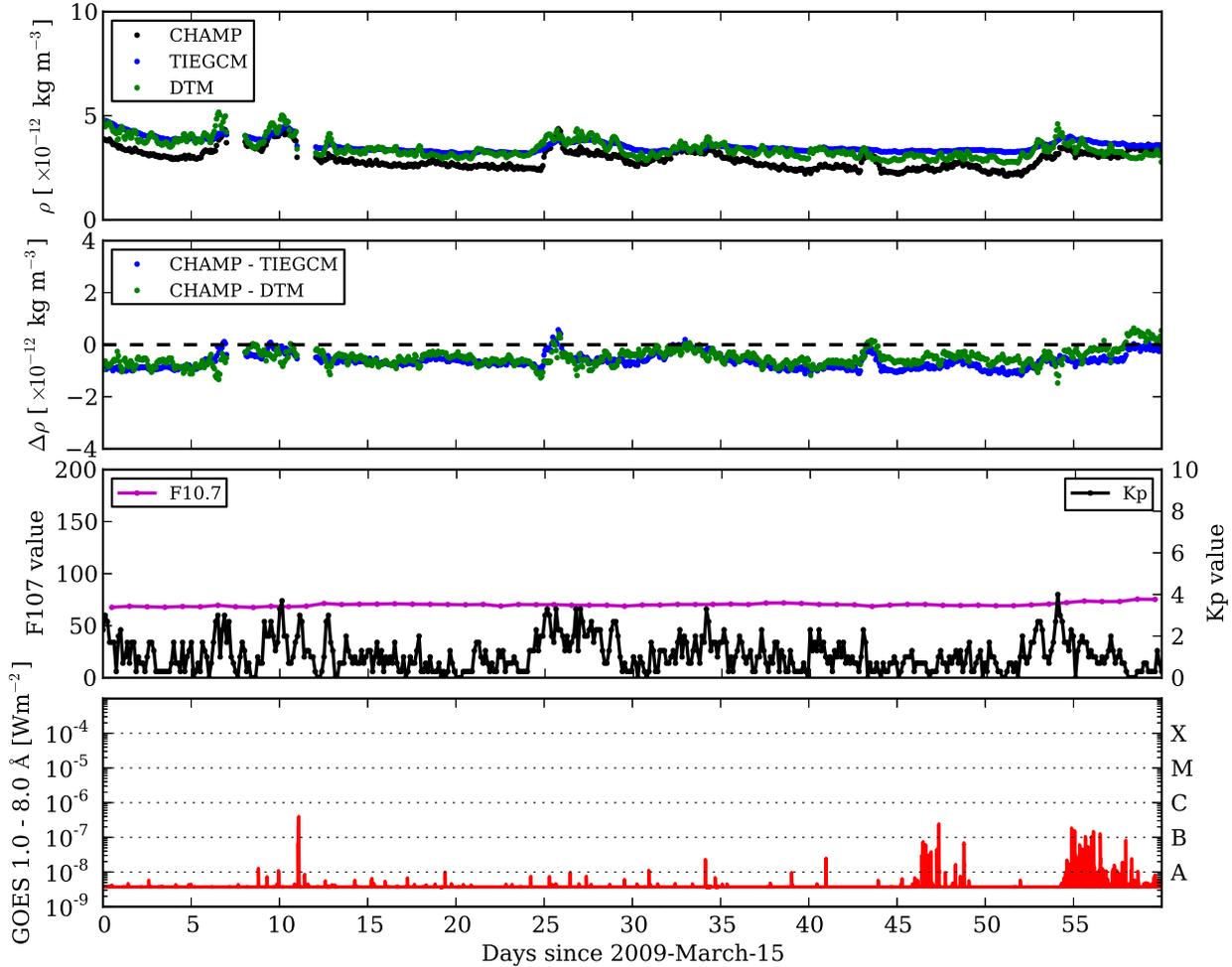}
\caption{Upper row: Orbit-averaged CHAMP (black), DTM (green), and TIEGCM (blue) densities for 2009 March 15 to 2009 May 13. Upper middle: Density difference between CHAMP observations and the two models. Lower middle row: F10.7 (magenta) and K$_\mathrm{p}$ (black) values used as inputs to the models. Lower row: GOES peak flux (red), where the dashed horizontal lines intersecting the right y-axis indicate flare class. The altitude of the CHAMP spacecraft during this period varied between $\sim$350--316~km.}
\label{figure_200903}
\end{figure*}

The results of the thermospheric model comparison for solar minimum are presented in Figure~\ref{figure_200903}. The lower middle row shows the F10.7 solar flux (magenta) and K$_\mathrm{p}$ indices (black) that were used as inputs to both models. With F10.7 virtually constant at $\sim$70, and K$_\mathrm{p}$ generally below $\sim$3, this can be considered a quiet period in terms of storms, and thus a good indicator of results at typical solar minimum conditions. In soft X-ray range, solar flares are classified as A-, B-, C-, M- or X- class according to peak flux measured near Earth by the Geostationary Operational Environmental Satellite (GOES) over  1--8 \AA~in Watts m$^{-2}$. Solar activity was extremely low throughout this period, with no solar flares greater than B-class, as indicated by the GOES flux plot in the lower row of Figure~\ref{figure_200903}. Two B-class flares occurred on 2009 March 26 (see lower row of Figure~\ref{figure_200903}), with no major active regions on the solar disk at the time. A larger number of A- and B- class flares occurred at the end of the period, again with no major active regions on the solar disk.

The upper row of Figure~\ref{figure_200903} shows the orbit-averaged density values of CHAMP (black), TIEGCM (blue), and DTM (green). The upper middle row shows the density difference between CHAMP and TIEGCM (blue), and CHAMP and DTM (green). The results indicate that DTM performs better at solar minimum, with its resulting densities closest to the actual observations from CHAMP.  However, TIEGCM also performs very well, with densities only slightly higher than DTM overall. The density values are relatively stable throughout the period. Both models mirror any gradual changes in density, with DTM mirroring rapid changes particularly well during solar minimum conditions.

The upper row of Figure~\ref{combined_short} shows the same results as Figure~\ref{figure_200903}, however for a shorter time period of 12 hours on 2009 March 15. This zoom-in reveals greater detail in the evolution of the densities, showing typical motions of the CHAMP spacecraft as it changes in altitude, latitude, and longitude. Both models mirror the CHAMP trend well, with DTM performing better than TIEGCM as the spacecraft crosses the night-side. It is clear from both figures that the models generally overestimate the density during solar minimum, with resulting model densities consistently higher than observations throughout the period of analysis. 

\begin{figure*}[t]
\centering
\noindent\includegraphics[width=\textwidth]{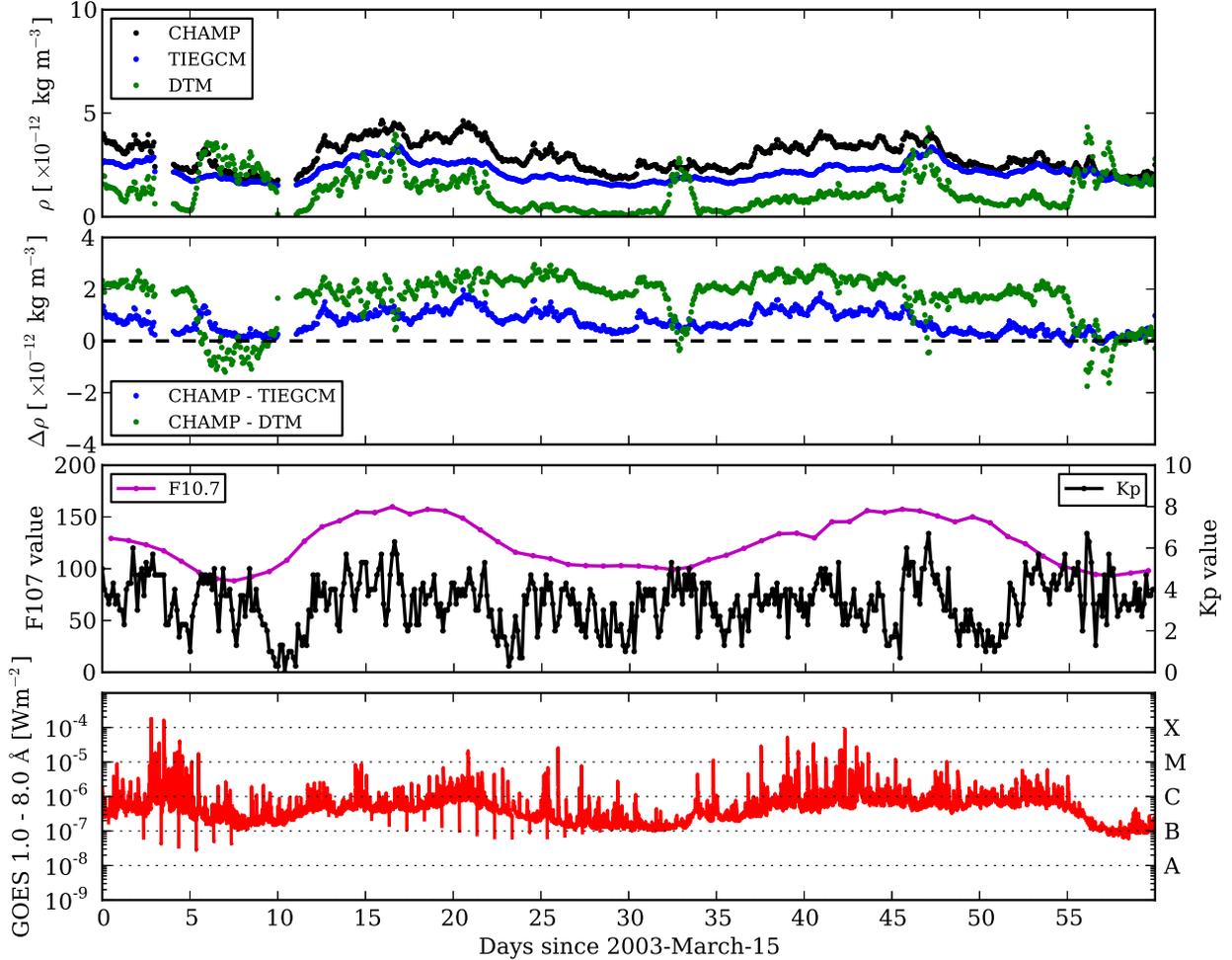}
\caption{Upper row: Orbit-averaged CHAMP (black), DTM (green), and TIEGCM (blue) densities for 2003 March 15 to 2003 May 13. Upper middle: Density difference between CHAMP observations and the two models. Lower middle row: F10.7 (magenta) and K$_\mathrm{p}$ (black) values used as inputs to the models. Lower row: GOES peak flux (red), where the dashed horizontal lines intersecting the right y-axis indicate flare class. The altitude of the CHAMP spacecraft during this period varied between $\sim$433--399~km.}
\label{figure_200303}
\end{figure*}

The same analysis was performed on CHAMP and both models at solar maximum, the results of which are shown in Figure~\ref{figure_200303}. It is clear from the lower middle row of the figure that there were more active conditions during this period, with F10.7 solar flux varying between $\sim$100--160, and K$_\mathrm{p}$ index reaching $\sim$7. Although no significantly large storms occurred during this time, the values are sufficiently high to contrast with the solar minimum results. Solar activity was much higher, with consistent flaring throughout the period, including multiple large flares (see lower row of Figure~\ref{figure_200303}). For example, a number of M- and X- class flares occurred between 2003 March 17--19 from a very complex active region just west of solar disk centre. This region was classified as $\beta\gamma\delta$ according to the modified Mount Wilson sunspot classification scheme \citep{brayloughhead64}. The impact of these eruptions is not noticeable from the F10.7 and K$_\mathrm{p}$ values at this time, with no significant Earth-directed coronal mass ejections associated with the flares.

As indicated by the orbit-averaged densities and difference shown in the upper and upper middle rows of Figure~\ref{figure_200303}, TIEGCM generally outperforms DTM at solar maximum. DTM does mirror increases in the CHAMP densities better than the physical model at the sharp increase seen on 2003 March 20 in CHAMP density measurements (likely related to the increased solar activity at this time). DTM also shows a sharp increase on April 30 that mirrors a smaller increase in CHAMP. TIEGCM densities do not change significantly on March 20, but do slightly increase on April 30. However, DTM also shows sharp increases/decreases in values during periods of increased K$_\mathrm{p}$ index that are not reflected in CHAMP or TIEGCM densities (e.g., April 16 and May 9). This suggests DTM is more sensitive to sharp changes in input parameter values than TIEGCM is. The upper row of Figure~\ref{combined_short} shows the solar maximum results over a shorter 12-hour period on 2003 March 15. This confirms that both models generally underestimate thermospheric density at solar maximum (in contrast to solar minimum results), and TIEGCM generally performs better during this period.

The results of analysis during the 2003 October and November storms are shown in Figure~\ref{figure_200310}. The significantly stormy conditions that occur towards the end of 2003 October are evident from GOES soft X-ray flux in the lower row, with multiple X-class flares, including an X17.2 flare on October 28. This resulted in a G5 level (http://www.swpc.noaa.gov/NOAAscales/) geomagnetic storm at Earth that lasted two days. The source regions of these large eruptions are discussed in detail by \citet{zhang03}. It is worth noting that the largest flare ever measured occurred on November 4, believed to be as large as X45 \citep{thomson04}. However, this event was not Earth-oriented and only resulted in high-latitude auroras. The sequence of events are reflected in the large K$_\mathrm{p}$ and F10.7 values shown in the lower middle row of Figure~\ref{figure_200310}. A second large peak in K$_\mathrm{p}$ occurs on November 20, however only M-class flares occurred during this period and F10.7 did not increase as much as previously.

The density values and differences shown in the upper and upper middle rows of Figure~\ref{figure_200310} reflect the particularly active conditions, with two large peaks in density at the end of October and November. Both models perform well in mirroring the observed increased densities, however they both underestimate the values again, particularly at the large peaks. TIEGCM more accurately follows the changes in CHAMP values, with DTM showing a more sporadic variance. Towards the end of the time period DTM values sharply rise, which is not reflected in the observed values. However, this rise does occur soon after an increase in K$_\mathrm{p}$. The zoomed-in plots in the lower row of Figure~\ref{combined_short} confirm TIEGCM being the more accurate model in this case, since it generally follows changes in CHAMP values better than DTM. Compared to results from 2003 March, it seems that TIEGCM reacts best to sharp rises in K$_\mathrm{p}$ values greater than $\sim 8$, and does not react as significantly to changes in more moderate values. \citet{carter14} found the most important source of variability in TIEGCM originates from the K$_\mathrm{p}$ index, so it is unsurprising that it the model results reflect sharp changes in this parameter.

GRACE observations were also available for two time periods at solar minimum (2009 March) and maximum (2003 October). The same analysis as above was performed using GRACE data for these periods, as well as using the data for further analysis of the previous results. The resulting densities from running TIEGCM with assimilation of CHAMP data were interpolated to the altitude, latitude, and longitude of GRACE at the time of the observation in order to gain a better insight into the overall impact of running TIEGCM with assimilation using such a small amount of data. This ensures no bias with interpolating CHAMP-assimilated TIEGCM results to CHAMP observations (and then comparing with same). 

Results from the solar minimum period beginning 2009 March 15 (upper and upper middle rows) and the stormy period beginning 2003 October 15 (lower middle and lower rows) are shown in Figure~\ref{grace_interp}, including DTM runs for comparison. DTM was run for the particular time, altitude, latitude, and longitude of GRACE as previously. Comparing the results for the solar minimum period in Figure~\ref{figure_200903}, TIEGCM performs similarly within the same error range, however DTM more significantly outperforms TIEGCM compared to previously. A discrepancy occurs on 2009 May 3 when both models show a slight increase compared to a drop in observed density values, however this is likely due to an issue with the GRACE measurements for that particular date rather than the model results. It is also interesting to note that the models underestimate the observed GRACE values during this period, while they overestimated CHAMP values previously. Comparing the results from the Hallowe'en storm period in Figure~\ref{grace_interp} to those shown in Figure~\ref{figure_200310}, this time TIEGCM outperforms DTM, as it did when compared to CHAMP observations. TIEGCM density values match GRACE observations closer than DTM, which has a large number of outliers during the stormy peaks at the end of 2003 October and 2003 November.

All results from the model comparisons are summarised in Table~\ref{table}, showing monthly mean percentage difference between measured satellite densities and the resulting densities, from various model runs. As found above, the table confirms that DTM performs best compared to CHAMP at solar minimum, followed closely by TIEGCM, with both models slightly overestimating the CHAMP density values. DTM densities are much closer to GRACE observations compared to TIEGCM during this period, with both models underestimating values at the higher satellite altitude. There is a clear difference during solar maximum, with TIEGCM being more accurate at these times, and both models underestimating the CHAMP density associated with higher activity. The sharp changes in density during the Hallowe'en storms result in values being smaller during this period, however TIEGCM again outperforms DTM here, most significantly when compared with the GRACE observations (as is clear from Figure~\ref{figure_200310}, DTM overestimates the density values at the peaks). 

The table also notes the differences between running TIEGCM with and without data assimilation, as it is worth determining how much of an improvement the assimilation system made to the physical model. Similar comparisons for DTM are presented in \citet{bruinsma12}. The results indicate that using the EnOI data assimilation technique has a small positive impact on the TIEGCM results, improving performance by $\sim$4\% overall when a 1-hour assimilation cycle is used. The data assimilation technique improves the results greater during periods of solar maximum than minimum, the largest improvement found during the Hallowe'en storms.

%------------------------------------

\section{Discussion and Conclusions}
\label{concl}

\begin{figure*}[t]
\centering
\noindent\includegraphics[width=\textwidth]{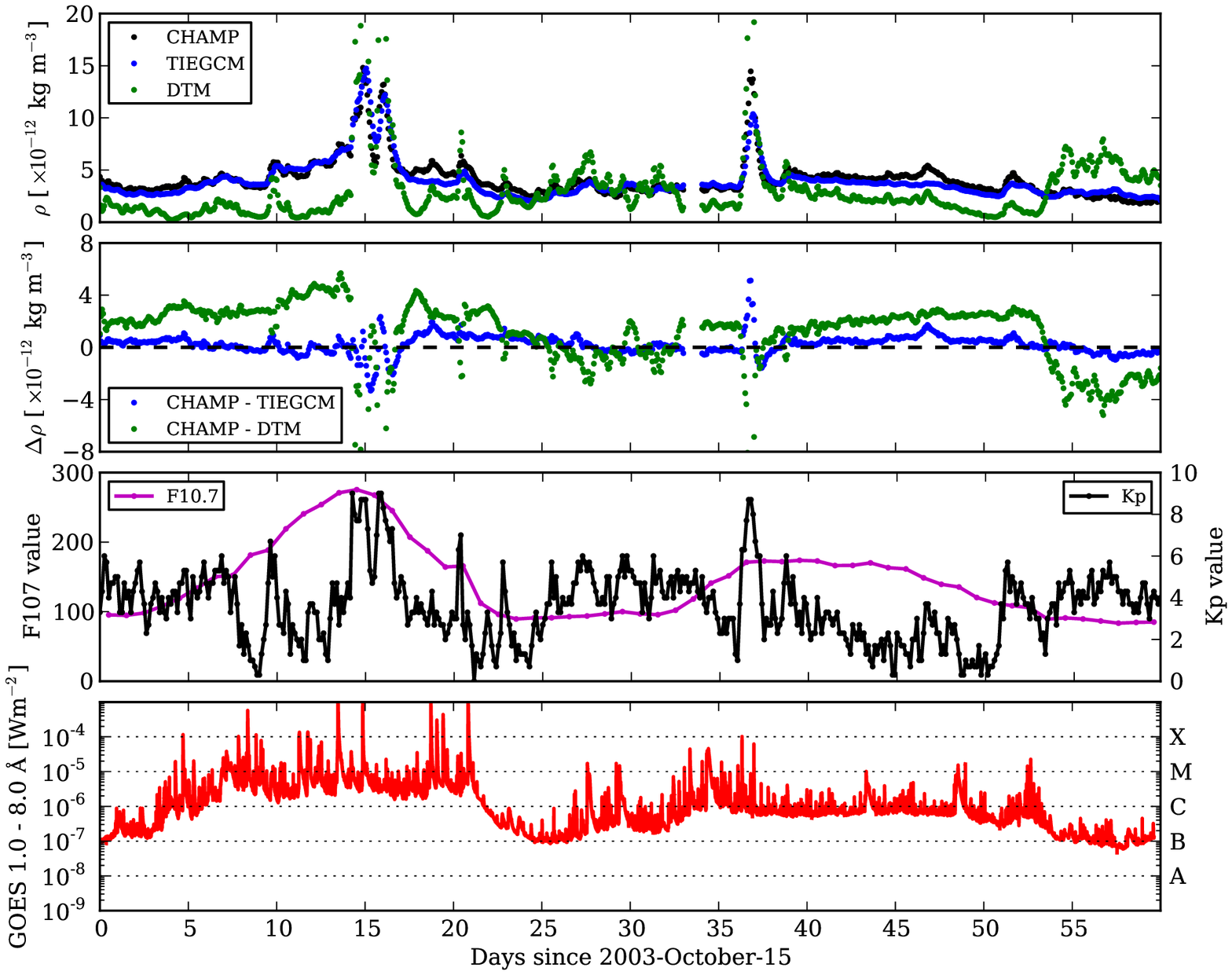}
\caption{Upper row: Orbit-averaged CHAMP (black), DTM (green), and TIEGCM (blue) densities for 2003 Ocotber 15 to 2003 December 13. Upper middle: Density difference between CHAMP observations and the two models. Lower middle row: F10.7 (magenta) and K$_\mathrm{p}$ (black) values used as inputs to the models. Lower row: GOES peak flux (red), where the dashed horizontal lines intersecting the right y-axis indicate flare class. The altitude of the CHAMP spacecraft during this period varied between $\sim$422--386~km.}
\label{figure_200310}
\end{figure*}

This paper has characterised the relative merit of two new modelling approaches with data assimilation capabilities developed during the ATMOP project; the physical model, TIEGCM, and the semi-empirical model, DTM. This has been done for 60-day periods for equinox at solar minimum (from 2009 March), solar maximum (from 2003 March), and the Hallowe'en 2003 storms (from 2003 October). Model results were validated against satellite data to investigate whether the improved physical model can outperform the semi-empirical model that is currently used operationally.

The solar minimum equinox results show that the models represent the thermospheric density well, with the DTM model being closest to CHAMP and GRACE observations, followed closely by TIEGCM. Interestingly, both models typically overestimate the CHAMP density, but underestimate the GRACE density. It is worth noting that satellite drag data studied by \citet{solomon11} indicates that the thermosphere was lower in density, and therefore cooler, during the protracted solar minimum period of 2007-2009 than at any other time in the past 47 years. Thus the overestimated values compared to CHAMP here may be representative of this particular `low' rather than representative of general solar minimum conditions - other periods would need to be examined for a more comprehensive picture. GRACE densities were taken from a higher altitude than the \citeauthor{solomon11} study.

\begin{table}[t]
\caption{Mean percentage difference between measured CHAMP (upper table) and GRACE (lower table) densities and resulting densities from various model runs. Values were calculated for the time periods beginning in March 2003, October 2003, and March 2009.}
\centering
\begin{tabular}{l l c c c c} 
\hline\hline
Validating  & Model Run & \multicolumn{3}{ c }{Mean \% Difference}  \\ 
\cline{3-5}
 Data&                    	& March `03 	& Oct. `03	& March `09      \\ 
 \hline
CHAMP &  TIEGCM  		& 30.09    	& 9.76 		& -26.86 	\\ 
&  TIEGCM (DA) 	& 25.50 	& 5.58 		& -23.35 	\\ 
&  DTM  			& 51.99 	& 19.27 	& -21.92 	\\ 
 \hline
 GRACE & TIEGCM                &  --          	& 10.66 	& 46.66       \\ 
  & TIEGCM (DA)     &  --           & 6.60 	        & 43.58       \\ 
   & DTM                  	&  --           & 43.33	        & 5.54         \\ 
    \hline\hline
\end{tabular}
\label{table}
\end{table}

The differing results for TIEGCM compared to these spacecraft observations may be related the anomalous constituent concentrations found at varying thermospheric heights during this deep solar minimum period. Previous work has found enhanced concentrations of helium, particularly in the winter hemisphere at $\sim$476km \citep[see][and references therein]{thayer12}, which is within the range of GRACE altitudes during the 2009 period studied here. Comparing CHAMP and GRACE observations in 2007 and 2008, \citeauthor{thayer12} found that altitude and latitude response in thermospheric mass density was influenced by the relative amount of helium (He) and oxygen (O) present. \citet{liu14} compared CHAMP and GRACE data to the semi-empirical NRLMISISE-00 model during the same 2008 period, finding wintertime helium concentrations exceeding model results by 30--70\%.

\begin{figure*}[t]
\centering
\noindent\includegraphics[width=\textwidth]{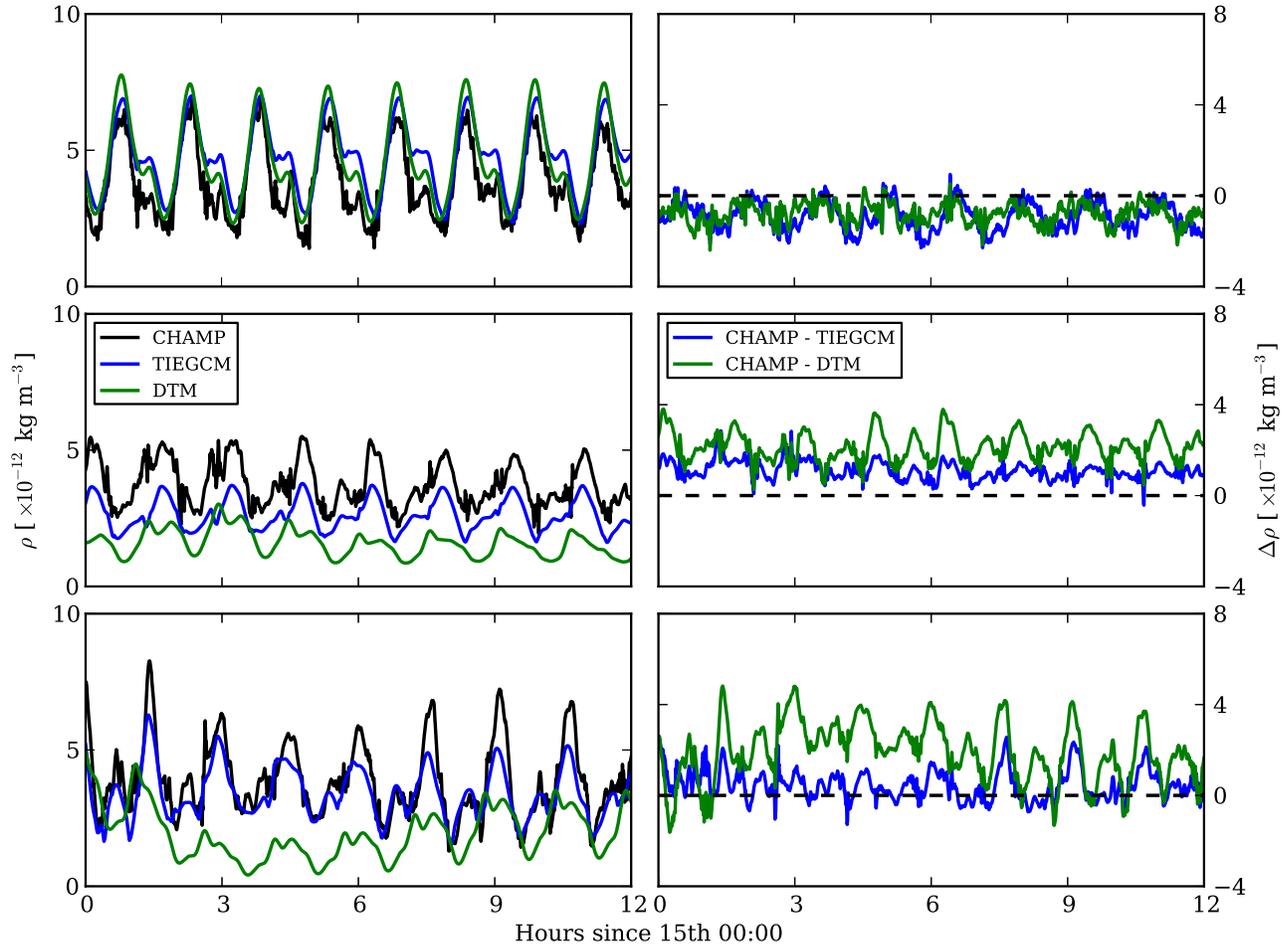}
\caption{Left column: CHAMP, DTM, and TIEGCM densities over 12 hours. Right column: Density difference between CHAMP observations and the two models. The upper to lower rows show results for 12 hours on 2009 March 15, 2003 March 15, and 2003 October 15, respectively. The altitude of the CHAMP spacecraft during these periods varied between $\sim$346--317~km, $\sim$432--401~km, and $\sim$422--391~km.}
\label{combined_short}
\end{figure*} 

\citeauthor{liu14} attributes the high concentrations to the extremely low He/O transition altitudes in the winter hemisphere in this cold, contracted thermosphere. This is a particularly important aspect to consider for forecasting purposes since the presence of He instead of O at the same temperature would act to increase the drag coefficient \citep{thayer12}. In this work DTM represents this differing concentration at GRACE altitudes better than the physical model, as is clear from Table~\ref{table}. DTM is trained on large observational data sets including the 2007--2008 period, which is more representative of the deep minimum conditions than data used for training in earlier versions of DTM and the NRLMISISE-00 model. The assimilation scheme used with TIEGCM only has a local impact at the lower CHAMP altitudes during the period being run.

\begin{figure*}[t]
\centering
\noindent\includegraphics[width=\textwidth]{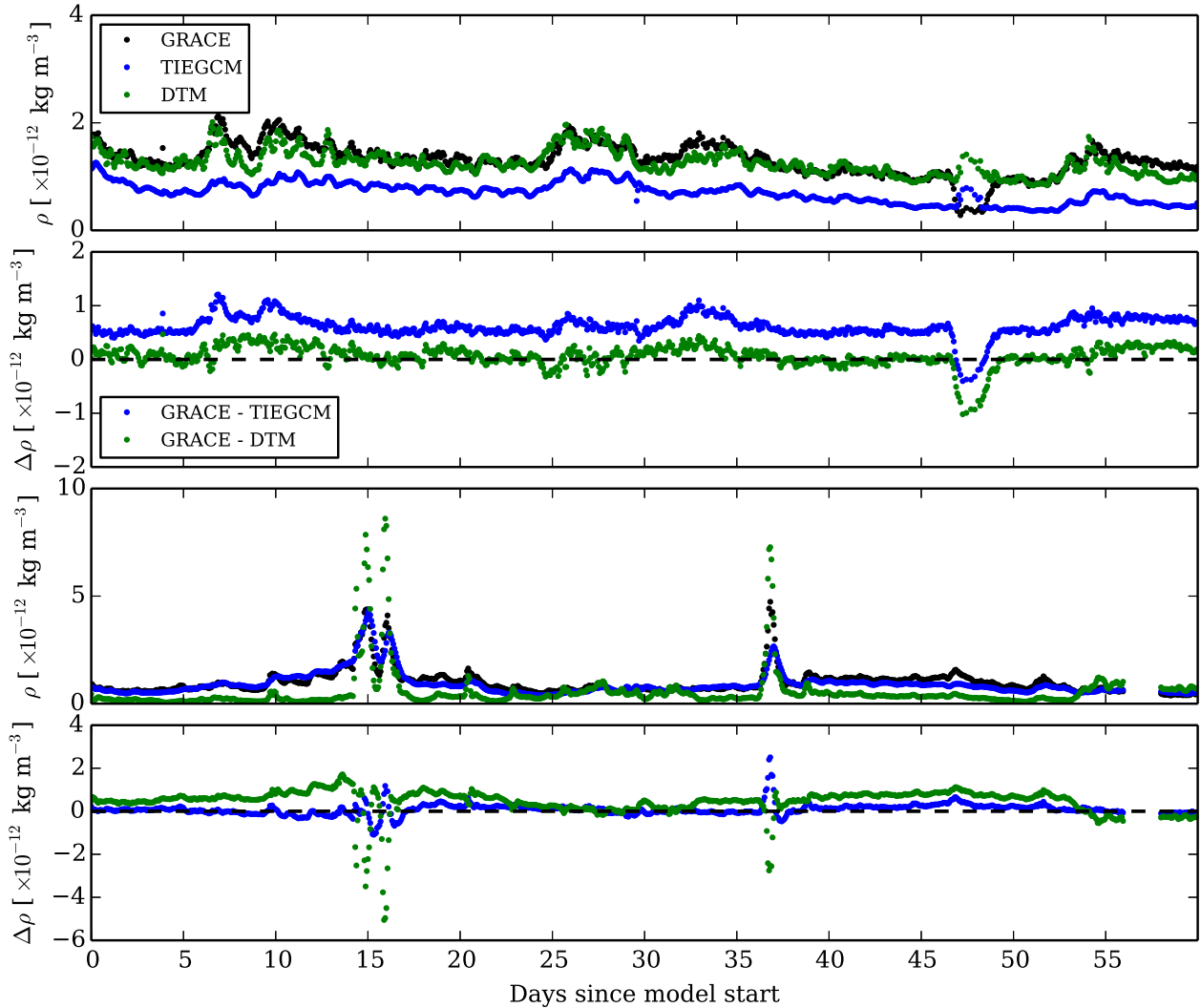}
\caption{Orbit-averaged DTM and TIEGCM with CHAMP DA densities compared with GRACE observations for the period beginning 2009 March 15 in the upper row, and 2003 October 15 in lower middle row. Density difference between GRACE and the two model runs are shown for the 2009 period in the upper middle row, and for the 2003 period in the lower row. The altitude of the GRACE spacecraft during the 2009 period varied between $\sim$506-454~km, and during the 2003 period varied between $\sim$522-466~km.}
\label{grace_interp}
\end{figure*}

It is clear from the results that while the semi-empirical model runs well during solar minimum, the physical model is more accurate when more stormy conditions are introduced. TIEGCM generally performs better during solar maximum conditions (Figure~\ref{figure_200903}), however DTM also mirrors changes in CHAMP densities well. Both models underestimate the observed density during the period beginning 2003 March. This is similar to the period of increased activity beginning 2003 October, where underestimations also exist (although TIEGCM densities match observations very well, as shown in Figures~\ref{figure_200310} and \ref{grace_interp}). \citet{sutton05} examines the geomagnetic impact of the Hallowe'en storms, and also finds the NRLMSISE-00 model to underestimate densities in comparison with CHAMP data during times of maximum geomagnetic activity. 

Since the Hallowe'en storms are such a well-studied event, our results from this period can be compared to previous studies. \citet{sutton05} find density measurements exhibit enhancements of 200--300\%. \citet{sutton06} uses GRACE and CHAMP data to observe the thermosphere neutral density response to the X-class flares on 2003 October 28 and November 4. \citeauthor{sutton06} finds an increase of $\sim$50--60\% in thermospheric density associated with the first flare at low to mid-latitudes, and a $\sim$35--45\% increase in response to the second flare. Figure 1 of \citet{qian12} shows the density response to the two flares using TIEGCM simulations as well as CHAMP observations. A $\sim$100--200\% density enhancement occurs, with storm response largest at high latitudes. These results compare well with the enhancements shown in Figure~\ref{figure_200310}, with increases of up to $\sim$200\% observed, and a slightly larger response found for the first flare compared to the second.

It is interesting to note there is a smaller enhancement observed in GRACE and interpolated model densities in Figure~\ref{grace_interp}, the observations obtained at a higher altitude (by $\sim$100~km) than the CHAMP observations. The storm-time responses examined here may also be affected by constituent composition varying with altitude, as mentioned previously for the solar minimum period studied. \citet{liu14} found the He enhancement causes differing perturbations of mass density with altitude after geomagnetic activity. In fact, these mass density perturbations are less enhanced in the winter hemisphere at GRACE altitudes ($\sim$25\%) than the lower CHAMP altitudes ($\sim$60\%) during the 2008 minimum. This could account for the smaller enhancements observed at GRACE altitudes here, however a more thorough investigation of latitudinal variation would be needed for confirmation.

Although the physical model clearly outperforms the semi-empirical model during stormy conditions, a number of improvements could still be made to the TIEGCM set-up used here. This work focuses on 2-month periods around solar equinox, and future work should also examine the solstice periods. However, for longer periods of study it must be considered that thermospheric neutral density and composition exhibits strong seasonal variation. Here the interaction with the lower atmosphere becomes important, however the standard version of TIEGCM used in this work only includes a simplified representation of the lower boundary (see Section~\ref{obs}). \citet{qian09} found that TIEGCM produces a better representation of the annual/semi-annual variation with gravity wave parameterisation and an eddy diffusion coefficient included at the lower boundary. The standard version of TIEGCM used in this work specifies an eddy diffusion coefficient as a constant at the lower boundary, which decreases exponentially with increasing pressure levels. \citeauthor{qian09} modified this globally uniform value to account for seasonal variation, representing the coefficient as a Fourier series with four harmonics per year. Research following on from the ATMOP project will include forcing the lower boundary of TIEGCM with a non-hydrostatic model of the lower atmosphere, which includes gravity wave parameterisation and non-migrating tides.

This work has presented validation periods at solar maximum and a particularly deep solar minimum. In order to gain a complete picture of how solar variability may be affecting the performance of the two models it may be useful to examine less deep solar minimum periods, as well as `solar medium' periods that lie midway through the ascending and descending phases of a solar cycle. Thermospheric models are known to be strongly influenced by external drivers, including geomagnetic and solar forcing besides boundary forcing. Much previous work has found that, alongside other improvements to the model, simply adjusting the F10.7 solar flux can improve results \citep{thayer12, matsuo13, liu14}. Future work incorporating both lower boundary improvements to TIEGCM and examining results over longer time periods will better highlight the possible impact that seasonal and solar variability may be having on the model densities.

It is clear from Figure~\ref{grace_interp} that interpolating the CHAMP-assimilated TIEGCM density results to higher satellite height did not have a major impact, with similar errors found to previous results. However, with such limited observations used during assimilation (one data point rather than a whole 2D map at each timestep) this is likely highlighting the accuracy of the model rather than saying too much about the assimilation procedure. As mentioned previously, the satellites are at slightly different local times, and enhanced He concentrations are found at the higher GRACE altitude particularly during the deep solar minimum in 2009. This likely introduces additional density errors that may counteract the positive impact of assimilating CHAMP data at these altitudes. At the lower altitudes, a positive but limited improvement has been obtained using data assimilation with a general circulation model, with an overall improvement of $\sim$4\% found. Including more relevant observations in the assimilation procedure would likely increase this percentage improvement.

Future work will also include improvements to the data assimilation techniques developed in the ATMOP project, particularly implementing incremental analysis updates (see Henley et al., manuscript in preparation, 2015, for further discussion). The EnOI method developed could be converted to an EnKF system without much difficulty. An EnOI system is similar to EnKF, however information from the observations which improve the main model does not feed back into the ensemble. Assessment of the other versions of DTM also developed as part of the ATMOP project could be undertaken using the same method as in this paper. DTM-2013 is supplemented by 2.5~years of GOCE satellite data, uses the 30~cm radio flux as a solar proxy, and 3-hour A$_{\mathrm{m}}$ index as geomagnetic proxy. To improve predictions to 3-days out, another method (DTM-nrt) was developed to predict temperature corrections to DTM based on a neural network. Recent research highlights that the DTM-nrt version of the model is more accurate than the DTM-2012 studied here for 24-hour forecasts \citep{choury13}. Comparisons to and assimilation of other types of observations would also be beneficial, for example Fabry-Per{\'o}t interferometer wind measurements, or total electron content. An increase in the availability of more real-time in-situ observations of the thermosphere will aid this work and space weather forecasting in general. Data from missions such as SWARM \citep{friis06} and Drag and Atmospheric Neutral Density Explorer \citep{pilinski08} may prove useful for this purpose.

With further improvements, the use of general circulation models in operational forecasting, in addition to empirical models currently used, is certainly plausible. Future work will allow near-real-time assimilation of thermospheric data into TIEGCM for forecasting.

%%% End of body of article:

%%%%%%%%%%%%%%%%%%%%%%%%%%%%%%%%
%% Optional Appendix goes here
%
% \appendix resets counters and redefines section heads
% but doesn't print anything.
% After typing \appendix
%
%\section{Here Is Appendix Title}
% will show
% Appendix A: Here Is Appendix Title
%
%%%%%%%%%%%%%%%%%%%%%%%%%%%%%%%%%%%%%%%%%%%%%%%%%%%%%%%%%%%%%%%%
%
% Optional Glossary or Notation section, goes here
%
%%%%%%%%%%%%%%
% Glossary is only allowed in Reviews of Geophysics
% \section*{Glossary}
% \paragraph{Term}
% Term Definition here
%
%%%%%%%%%%%%%%
% Notation -- End each entry with a period.
% \begin{notation}
% Term & definition.\\
% Second term & second definition.\\
% \end{notation}
%%%%%%%%%%%%%%%%%%%%%%%%%%%%%%%%%%%%%%%%%%%%%%%%%%%%%%%%%%%%%%%%
%
%  ACKNOWLEDGMENTS

\begin{acknowledgments}
This work was supported by the European Framework Package 7 Advanced Thermosphere Modelling for Orbit Prediction project (Work Package 5.6). Data from CHAMP and GRACE missions are made available to the community by the Information Systems and Data Center at GFZ (http://isdc.gfz-potsdam.de). Model code for TIEGCM and DTM are made freely available to the community by NCAR (http://www.hao.ucar.edu/modeling/tgcm/) and the ATMOP project (http://www.atmop.eu) respectively. The authors thank the anonymous referees for their constructive suggestions to improve the manuscript.

\copyright~British Crown Copyright 2015, the Met Office.
\end{acknowledgments}

%% ------------------------------------------------------------------------ %%
%%  REFERENCE LIST AND TEXT CITATIONS
%
% Either type in your references using
% \begin{thebibliography}{}
% \bibitem{}
% Text
% \end{thebibliography}
%
% Or,
%
% If you use BiBTeX for your references, please use the agufull08.bst file (available at % ftp://ftp.agu.org/journals/latex/journals/Manuscript-Preparation/) to produce your .bbl
% file and copy the contents into your paper here.
%
% Follow these steps:
% 1. Run LaTeX on your LaTeX file.
%
% 2. Make sure the bibliography style appears as \bibliographystyle{agufull08}. Run BiBTeX on your LaTeX
% file.
%
% 3. Open the new .bbl file containing the reference list and
%   copy all the contents into your LaTeX file here.
%
% 4. Comment out the old \bibliographystyle and \bibliography commands.
%
% 5. Run LaTeX on your new file before submitting.
%
% AGU does not want a .bib or a .bbl file. Please copy in the contents of your .bbl file here.

%\bibliographystyle{agufull08}
%\bibliography{bibliography}

%\begin{thebibliography}{}

% \end{thebibliography}
%
% Please use ONLY \citet and \citep for reference citations.
% DO NOT use other cite commands (e.g., \cite, \citeyear, \nocite, \citealp, etc.).

%% ------------------------------------------------------------------------ %%
%
%  END ARTICLE
%
%% ------------------------------------------------------------------------ %%
\end{article}
%
%
%% Enter Figures and Tables here:
%
% DO NOT USE \psfrag or \subfigure commands.
%
% Figure captions go below the figure.
% Table titles go above tables; all other caption information
%  should be placed in footnotes below the table.
%
%----------------
% EXAMPLE FIGURE
%
 %\begin{figure}
 %\noindent\includegraphics[width=20pc]{samplefigure.eps}
 %\caption{Caption text here}
 %\label{figure_label}
 %\end{figure}
%

% ---------------
% EXAMPLE TABLE
%
%\begin{table}
%\caption{Time of the Transition Between Phase 1 and Phase 2\tablenotemark{a}}
%\centering
%\begin{tabular}{l c}
%\hline
% Run  & Time (min)  \\
%\hline
%  $l1$  & 260   \\
%  $l2$  & 300   \\
%  $l3$  & 340   \\
%  $h1$  & 270   \\
%  $h2$  & 250   \\
%  $h3$  & 380   \\
%  $r1$  & 370   \\
%  $r2$  & 390   \\
%\hline
%\end{tabular}
%\tablenotetext{a}{Footnote text here.}
%\end{table}

% See below for how to make sideways figures or tables.

\end{document}